\documentclass[10pt,final,twocolumn,journal]{IEEEtran}
\usepackage[noadjust]{cite}
\usepackage{cite}
\usepackage{graphicx}
\usepackage{caption}
\usepackage{subcaption}
\usepackage{authblk}
\usepackage{xcolor}
\usepackage{amsmath}
\usepackage{array}
\usepackage[font=small,labelfont=bf]{caption}

\begin{document}
\title{Dramatic Impact of Dimensionality on the Electrostatics of PN Junctions}
\author{Hesameddin Ilatikhameneh, Tarek Ameen, Fan Chen, Harshad Sahasrabudhe, Gerhard Klimeck, Rajib Rahman \vspace{-6.5ex}
\thanks{This work was supported in part by the Center for Low Energy Systems Technology (LEAST), one of six centers of STARnet, a Semiconductor Research Corporation program sponsored by MARCO and DARPA.}
\thanks{The authors are with the Department of Electrical and Computer Engineering, Purdue University, West Lafayette, IN, 47907 USA e-mail: hesam.ilati2@gmail.com.}
}
\maketitle
\setlength{\textfloatsep}{12pt} 
\setlength{\belowdisplayskip}{1.6pt} 
\setlength{\belowdisplayshortskip}{1.6pt}
\setlength{\abovedisplayskip}{1.6pt} 
\setlength{\abovedisplayshortskip}{1.6pt}
\setlength{\belowcaptionskip}{-12pt}
\vspace{-1.0\baselineskip}
\begin{abstract}
Low dimensional material systems provide a unique set of properties useful for solid-state devices. The building block of these devices is the PN junction. In this work, we present a dramatic difference in the electrostatics of PN junctions in lower dimensional systems, as against the well understood three dimensional systems. Reducing the dimensionality increases the depletion width significantly. We propose a novel method to derive analytic equations in 2D and 1D that considers the impact of neutral regions. The analytical results show an excellent match with both the experimental measurements and numerical simulations. The square root dependence of the depletion width on the ratio of dielectric constant and doping in 3D changes to a linear and exponential dependence for 2D and 1D respectively. This higher sensitivity of 1D PN junctions to its control parameters can be used towards new sensors. 
\end{abstract}
\vspace{-0.5\baselineskip}
\section{Introduction}
Low dimensional materials have become the building blocks of nanotechnology. The fundamental properties of 1D and 2D materials offer a new toolbox for electronic \cite{Geim}, thermoelectric \cite{Dres}, and optoelectronic  \cite{MoS2_photo} applications; like the high mobility graphene, carbon nanotube, and III-V nanowires for transistors \cite{FrankGR, CNT, Fan1, Fiori2, Fan2, MoS2, Mehdi}, transition metal dichalcogenides and phosphorene for tunnel transistors \cite{Hesam1, Sarkar, Hesam3, Fiori_Nature, Hesam2, Tarek1, Hesam4}, high ZT nanoribbon and nanowire thermoelectrics \cite{Thermo, Si_Thermo, Si_Thermo2} and 2D photodetectors \cite{MoS2_photo2, Tao}. Almost all electronic and optical devices are based on PN junctions. Hence, the knowledge of the electrostatics of PN junctions is critical in predictive device design. For example, the tunneling current in tunnel diodes \cite{Kane1, SB1, Analytic2} and tunnel FETs \cite{App1, Ionescu, Sub10, Analytic1} depends exponentially on the depletion width. Low dimensional material systems, however, present significant differences compared to their 3D counterparts, as detailed in the next section. 

\textcolor{black}{
Although the impact of the dimensionality of PN junctions on their electrostatics has usually been ignored in literature, there exist experiments supporting this claim. In the case of 2D materials, Reuter et al. have measured depletion widths (W$_D$) of a 2D PN junction as a function of reverse bias using optical techniques and have shown the difference in behavior of W$_D$ of 2D junctions as against their 3D counterparts \cite{Exp2D}. More recently, Zheng et al. have used Kelvin Probe Force Microscopy (KPFM) technique to show that the potential profile itself changes significantly in low dimensions \cite{Changxi}. 
}

\textcolor{black}{
Several theoretical models have been developed to explain some of these anomalous behavior in electrostatics of low dimensional systems. For 1D systems, Leonard and Tersoff at IBM noticed the new length scales in nanotubes and provided approximate analytic equations for depletion width \cite{Leonard}. In 2D, Achoyan et al. have derived equations for a 2D PN junction with zero thickness \cite{2d_theo_1}. Later on, Gharekhanlou et al. \cite{Khor1, Khor2} and Yu et al. \cite{2d_theo_2} have used different methods to model 2D junctions. However, in all of the above models, the thickness of the PN junction has been ignored and there is no validation against numerical simulations. These issues have been addressed in this paper.  
}

\textcolor{black}{
In this work, a novel approach is proposed for analytic modeling of PN junctions. The equations governing the potential profiles and depletion width of 3D, 2D and 1D systems are presented. Experiments on PN junctions demonstrate an enormous impact of dimensionality on the electrostatics \cite{Changxi, Exp2D, Exp1D}. The rapid progress in doping thin flakes of 2D materials and 1D nanowires over the last two decades \cite{Doping1, Doping2, Doping3} and its role in low dimensional MOSFETs \cite{Mehdi1, Mehdi2, Mehdi3} and tunnel transistors \cite{Hesam1, Sarkar, Hesam3, Fiori_Nature, Hesam2, Tarek1, Hesam4} call for a thorough analytical understanding of the low dimensional PN junctions, well validated against numerical simulations.
}

\textcolor{black}{
In the following sections, the numerical simulation method is discussed first. A novel approach to obtain the analytic equations, including the impact of neutral regions, is introduced next. The results of the analytical model are then compared with experimental measurements and numerical simulations. Finally, the approximate equations for depletion width of PN junctions in diffenret dimensions are presented. 
}
\begin{figure}[!t]
        \centering
        \begin{subfigure}[b]{0.49\textwidth}
               \includegraphics[width=\textwidth]{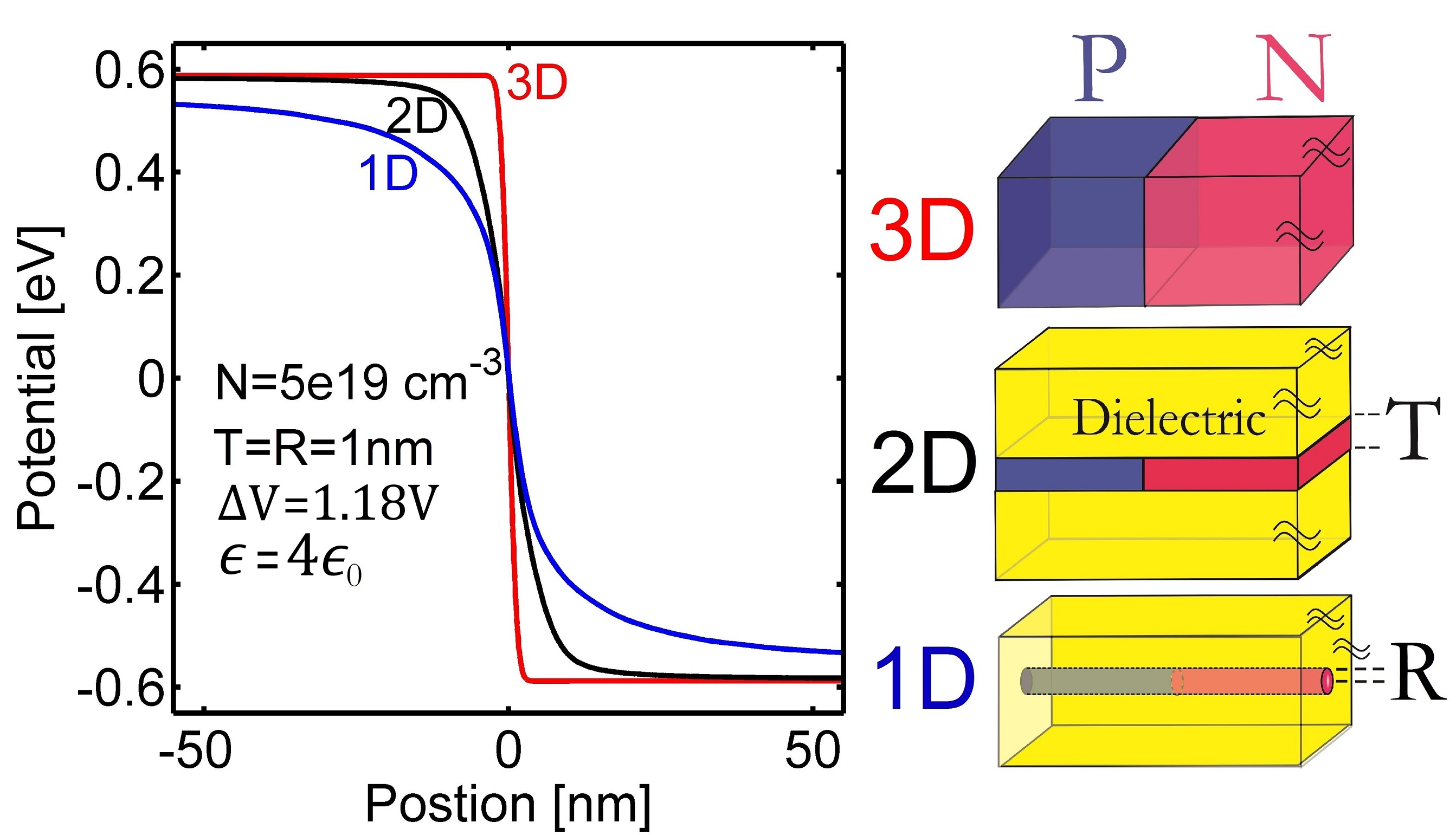}
                \label{fig:Same_EoT}
        \end{subfigure}%
        \vspace{-1.5\baselineskip}
        \caption{ \textbf{Potential profile of \textcolor{black}{abrupt} PN junctions in 3D, 2D, and 1D.} }\label{fig:Fig1}
\end{figure} 
\section{Numerical simulation method} 
\textcolor{black}{
The potential profile in 2D and 1D PN junctions, as shown in Fig. 1, are obtained by self-consistently solving Poisson equation with the drift-diffusion equation using the finite element method in NEMO5 \cite{N5, N5_2}. The carrier densities are obtained from Fermi integrals based on the difference between Fermi level and band edges. It is important to consider a large enough dielectric environment in the simulations (i.e. dielectric thickness larger than depletion width) since a zero electric field boundary condition is used at the dielectric boundaries of the simulation domain. If these boundaries are close to the junction, they can artificially impact the fringing field pattern and the potential profile within semiconductor. Therefore, it is critical to make the simulation domain large enough so that boundaries do not affect the results.
}

\textcolor{black}{
For the sake of simplicity, it is assumed here that the P and N doping levels are both equal to $N$ and the dielectric constants of channel and surrounding dielectric equal $\epsilon$. $\Delta V$ is the potential drop across the PN junction which equals $V_{bi}-V_{a}$ where $V_{bi}$ is the built-in potential and $V_{a}$ is the applied bias. 
}

Fig. 1 shows the simulated potential profile of PN junctions, in different dimensions, with the same doping of 5e19 cm$^{-3}$ using finite-element method. It is clear that the well known 3D electrostatics does not apply to low dimensional junctions \textcolor{black}{and dimensionality of material  has a significant impact on the depletion width and potential profile.} 

\section{Analytical model} %
In this section, the analytic method to obtain exact analytical equations for 3D, 2D and 1D PN junctions is introduced. 
\textcolor{black}{The depletion and neutral regions are shown in Fig. 2a. Free carriers in the neutral regions play an important role in screening the electric field. In the conventional depletion approximation, the electric field in neutral regions is zero. This zero field boundary condition, at the interface of neutral and depletion regions, can be captured by introducing image charges. Ignoring these image charges and considering only the E-field due to the charges in the depletion region leads to wrong results such as large electric fields at the boundaries of depletion region and non-monotonic potential profile in 2D and 1D \cite{Khor1, Khor2}}.
\begin{figure}[!t]
        \centering
        \begin{subfigure}[t]{0.47\textwidth}
               \includegraphics[width=\textwidth]{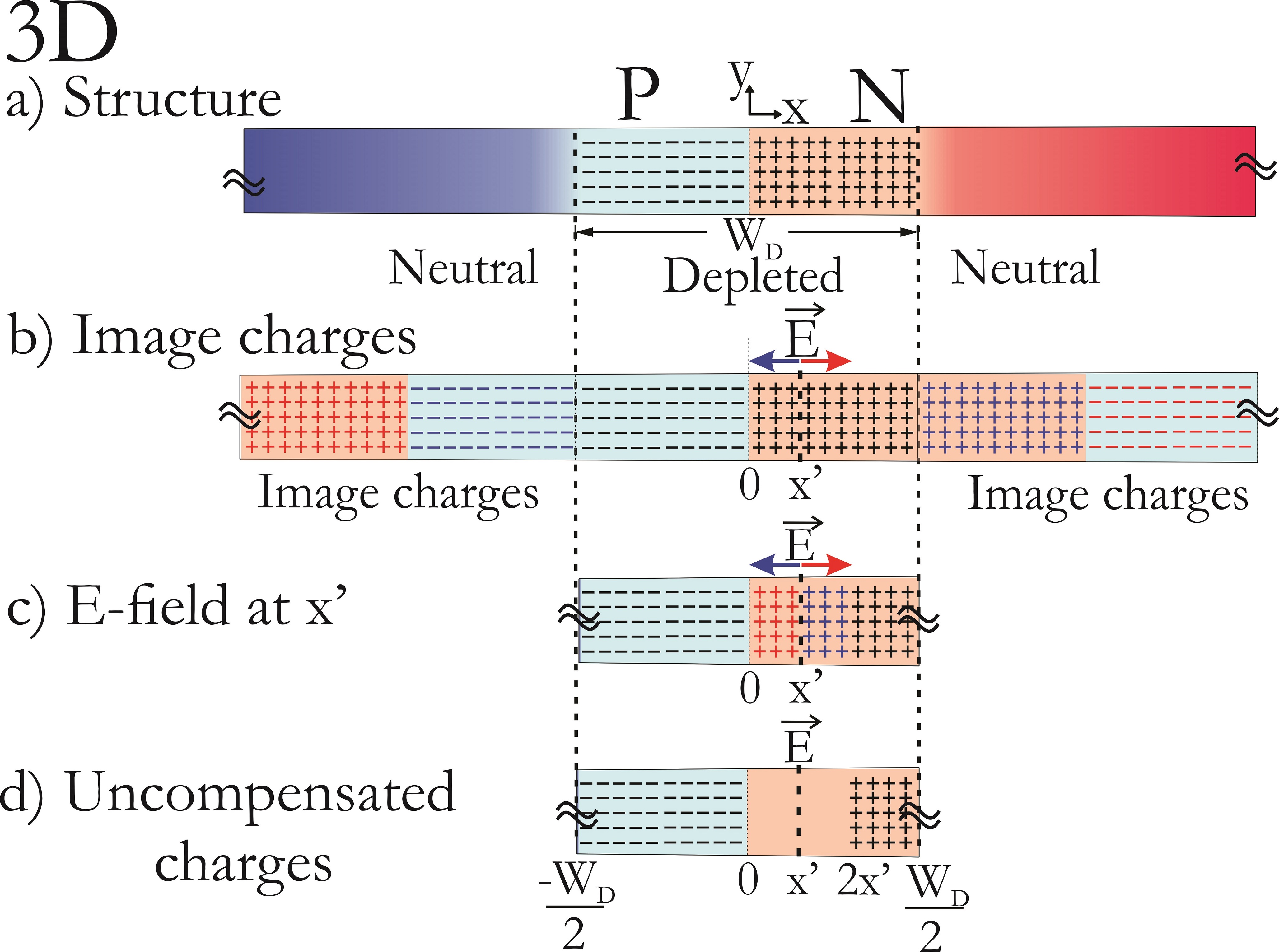}
        \end{subfigure}%
        \caption{ \textbf{Illustration of the analysis method in a 3D PN junction: a) neutral and depletion regions, b) image charges, c) E-field evaluation at position x', d) uncompensated charges contributing to the net electric field at x'.}}\label{fig:Fig2}
\end{figure}
\textcolor{black}{The steps involved in determining the analytic equation for potential are presented in Figs. 2 and 3: 1) Place image charges to enforce the zero E-field boundary condition at the interfaces of neutral and depletion regions. 2) Find the E-field at position $x'$. 3) Integrate E-field to obtain the potential. The 2nd step for evaluation of E-field involves the following: a) Removing charges whose E-fields cancel out. After these cancellations, a negative block of charge and a positive block of charge remain. b) E-field evaluation at position $x'$ due two remaining blocks of charge. For a symmetric case, it is sufficient to evaluate E-field due to one block and multiply it by 2.}

\textcolor{black}{
To demonstrate the new approach, we start with the well known 3D case. The potential in 3D case can be easily obtained by solving 1D differential equation but our purpose here is to introduce the method in the simplest case. Fig. 2b shows the equivalent structure where neutral regions are replaced by image charges to enforce zero E-field at the boundaries of the depletion region. Now that all charges are in place, the E-field at position $x'$ should be evaluated. First, the contribution of image charges to E-field is shown in Fig. 2b. The image charges inducing positive E-field at $x'$ are colored red and the opposite are blue.  Since E-field due to an infinite plane of charge is independent of the distance from the plane, the positive and negative E-fields due to the \textbf{image charges in 3D} cancel each other out. Moreover, as shown in Fig. 2c, the field due to charges from 0 to $x'$ compensates the ones from $x'$ to $2x'$. Fig. 2d shows all the charges which contribute to $E_{net}$, the net electric field at position $x'$. The field from the block of negative charges is proportional to their net charge and equals $-\frac{qN}{2\epsilon}W_D$, whereas the field due to positive block equals $\frac{qN}{2\epsilon}(2x'-W_D)$. Hence the net field equals to:  }  
\begin{equation}
\label{eq:opt1}
E_{net}^{3D}(x') = \frac{qN}{\epsilon} \left(x'- \frac{W_D}{2} \right),~ 0 < x' < \frac{W_D}{2}
\end{equation}
Integrating electric field gives the potential. Considering the reference potential to be the value at $x$=0 results in:
\begin{equation}
\label{eq:V3D}
V^{3D}(x) = - \int_0^x E(x') dx' = \frac{qN}{2 \epsilon} x \left( W_D - x \right)
\end{equation}
As expected, the potential profile has a parabolic form in the depleted region. From (\ref{eq:V3D}), $W_D$ can be derived: 
\begin{equation}
\label{eq:W3D}
W_D^{3D} = 2 \sqrt{ \frac{\Delta V \epsilon}{q N} }
\end{equation}
\begin{figure}[!t]
        \centering
        \begin{subfigure}[b]{0.47\textwidth}
               \includegraphics[width=\textwidth]{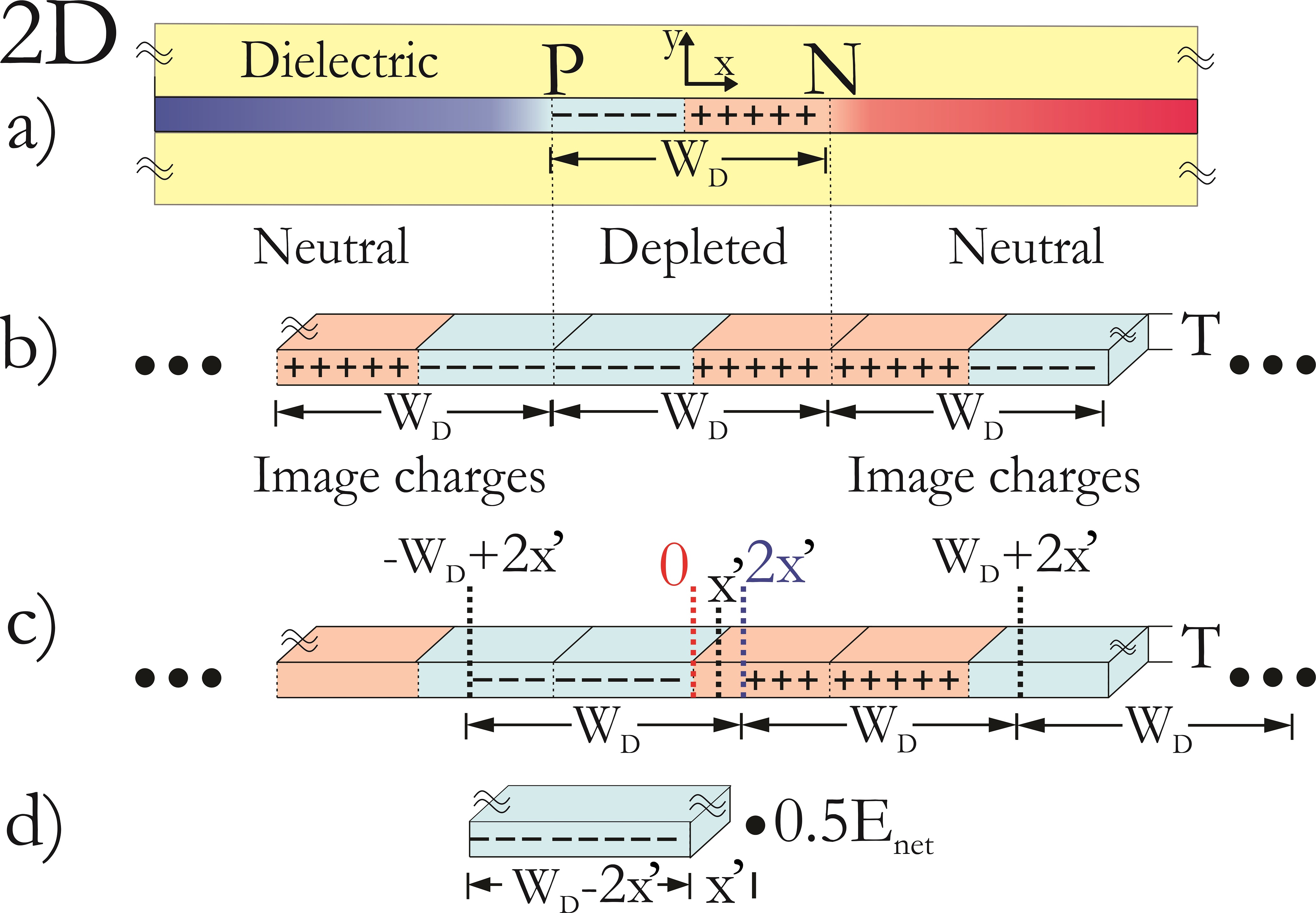}
                \label{fig:Same_EoT}
        \end{subfigure}%
        \caption{\textbf{Illustration of the analysis method in a 2D PN junction similar to Fig. 2.}}\label{fig:Fig3}
\end{figure}
\textcolor{black}{
Hence, we recover the well known electrostatics of bulk PN junctions. The same procedure can be followed for 2D junctions to find the potential profile as shown in Fig. 3. However, it is important to note that the net electric field due to image charges in 2D and 1D junctions is not zero, unlike in 3D, since the field from a finite block of charge now depends on the distance. The regions of image charges with positive and negative blocks of charge behave similar to electric dipoles with a finite net field. To the first order, the nearest neighbor block of image charges are considered first. The impact of charges further away are discussed later.} Fig. 3c shows the uncompensated charges for a 2D junction. Finally, the net electric field at a distance $x'$ is obtained from a 2D block of charge with length $W_D - 2x'$ and thickness $T$ and multiplied with a factor of 2 as shown in Fig. 3d:
\begin{figure}[!b]
        \centering
        \begin{subfigure}[b]{0.4\textwidth}
               \includegraphics[width=\textwidth]{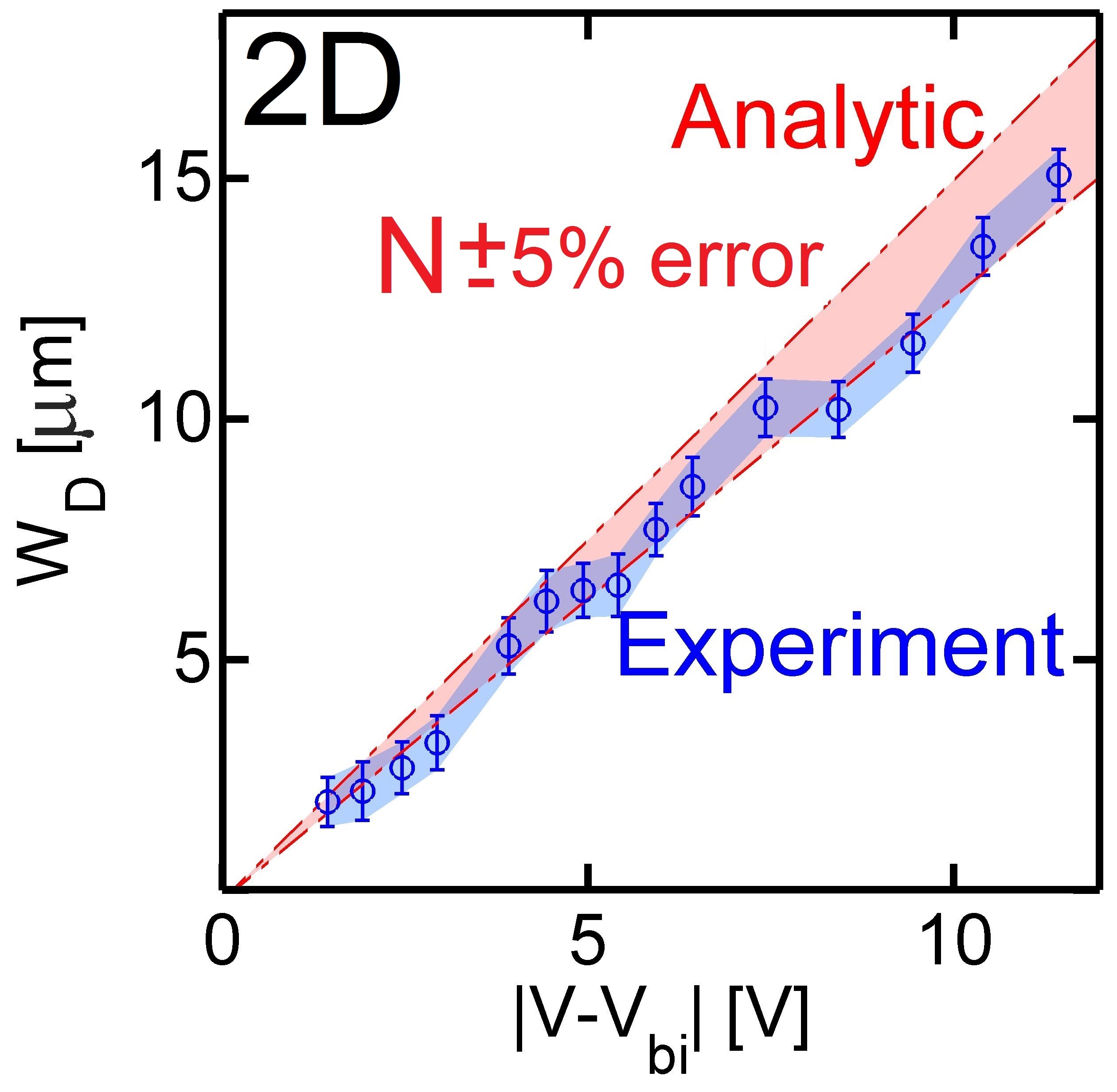}
                \label{fig:Same_EoT}
        \end{subfigure}%
        \vspace{-1.5\baselineskip}                    
        \caption{\textbf{Comparison between analytical and experimental results \textcolor{black}{\cite{Exp2D}} of a 2D junction with $\pm 5 \%$ tolerance in measured N.}}\label{fig:Fig4}
\end{figure}    
\begin{multline}
E_{net}^{2D}(x') =   \frac{-E_0}{2} ln\left( \frac{4(W_D-x')^2 + T^2}{4x'^2 + T^2} \right) \\
+ 2 E_0 \left( x'~atan\left(\frac{T}{2x'}\right) - (x' - W_D) ~ atan\left(\frac{T}{2(x' - W_D)}\right) \right)
\label{eq:opt1}
\end{multline}
where $E_0$ is defined as
\begin{equation} 
E_0 = \frac{q N}{\pi \epsilon}
\label{eq:opt1}
\end{equation}
To simplify the solution, the flake thickness can be assumed to be small compared with the depletion width ($T \ll W_D$ ) which leads to:
\begin{equation} 
E_{net}^{2D} \approx -E_0 ( ln(W_D - x') - ln(x') )
\label{eq:E2D}
\end{equation}
The potential is obtained from integrating electric field:
\begin{equation}
\label{eq:V2D}
\frac{V^{2D}(x)}{E_0} \approx  (x - W_D) ln(W_D -x)+W_D ln(W_D)-x~ln(x) 
\end{equation}
Evaluating (\ref{eq:V2D}) at $x$ = $W_D$/2 gives $W_D$ 
\begin{equation}
\label{eq:W2D}
W_D^{2D} \approx \frac{\Delta V}{ln(4) ~ E_0}
\end{equation}
Unlike in 3D, $W_D$ has a linear dependence on both $\Delta V$ and $\epsilon$ and is inversely proportional to $N$. To validate this significant difference from conventional PN junction behavior, the analytical results are compared with the experimental measurements of a PN junction made from 2DEGs (2 Dimensional Electron Gases) \textcolor{black}{\cite{Exp2D}}. Figure 4 shows a good match between the analytic calculations and \textcolor{black}{optical measurements \textcolor{black}{\cite{Exp2D}} of $W_D$ in 2D}.  
\begin{figure}[!t]
        \centering
        \begin{subfigure}[b]{0.48\textwidth}
               \includegraphics[width=\textwidth]{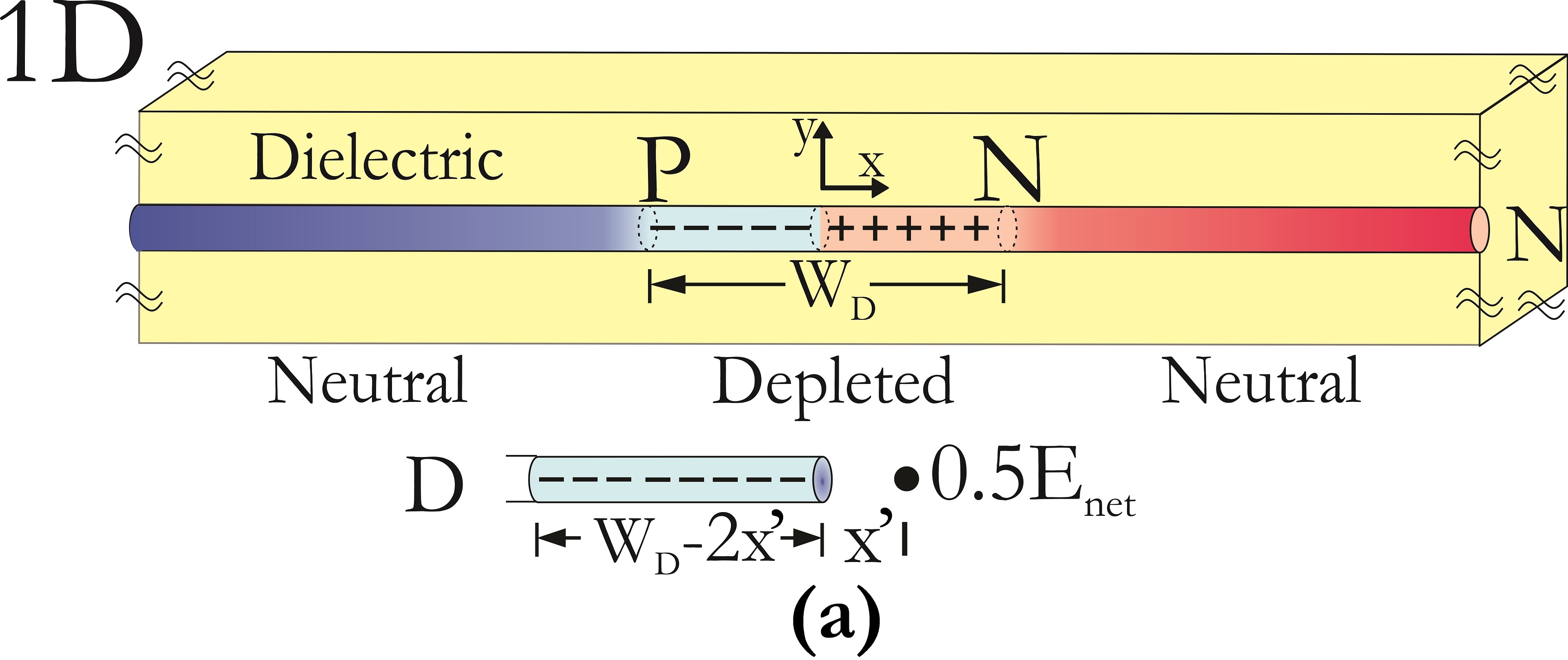} 
                \vspace{-0.8\baselineskip}                
                \label{fig:Same_EoT}
        \end{subfigure}%
        \quad
        \begin{subfigure}[b]{0.48\textwidth}
               \includegraphics[width=\textwidth]{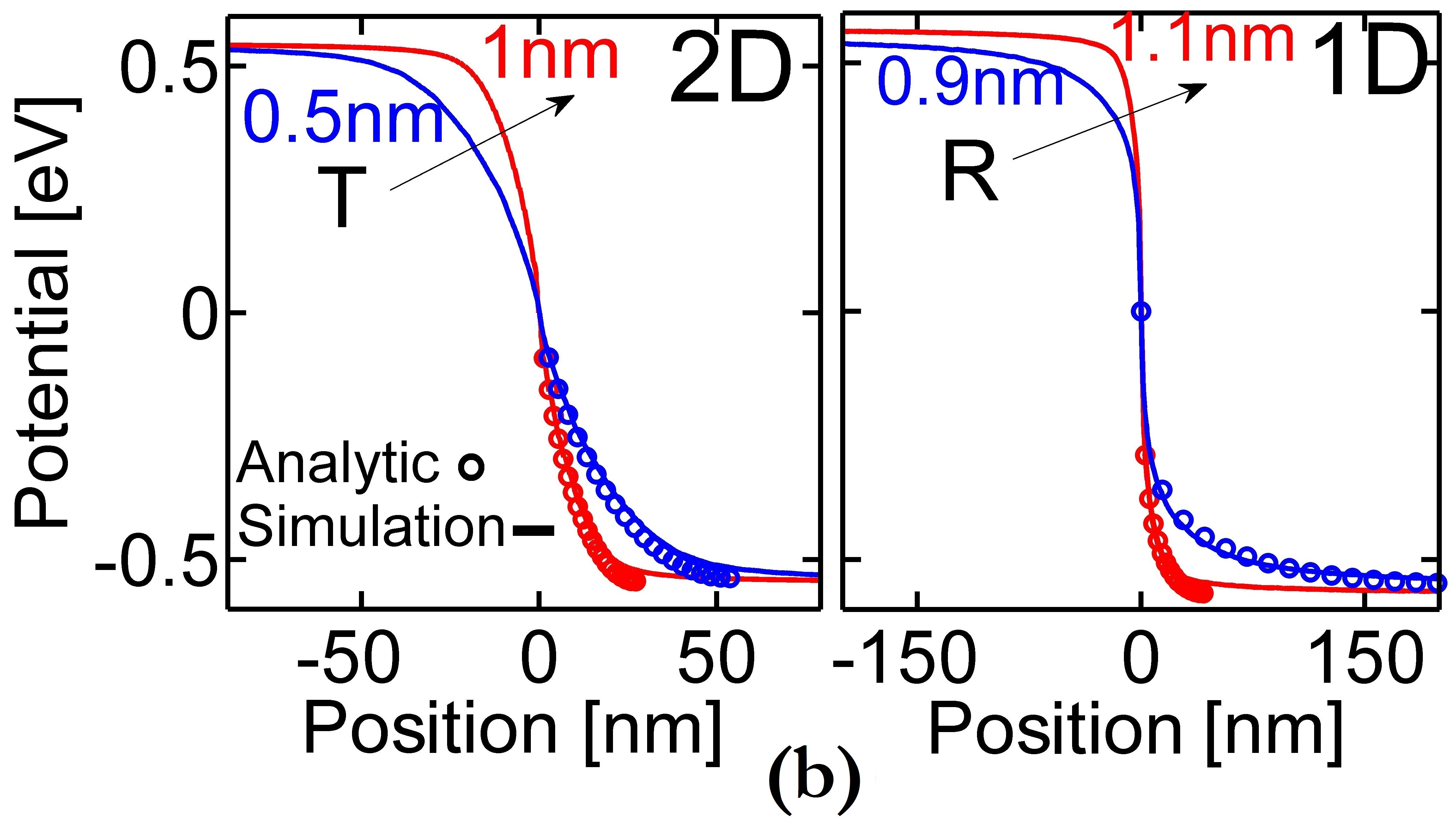} 
               \vspace{-1.5\baselineskip}
                \label{fig:Same_EoT}
        \end{subfigure}%
        \caption{\textbf{a) Structure and uncompensated charges contributing to $E_{net}$ in 1D PN junction. b) Simulated (lines) and analytic (circles) potential profile of 2D (left) and 1D (right) junctions with different thicknesses.} }\label{fig:Fig5}
\end{figure}

The analysis of 1D PN junction is similar to that of 2D. The only difference is that $E_{net}$ is due to a cylinder of charge with a length of $W_D - 2x'$ and diameter $D$ at a distance $x'$ (Fig. 5a):
\begin{equation}
E_{net}^{1D}(x') =  -\frac{q N}{\epsilon}  \left( (W_D - 2x') - L_{W_D - x'} + L_x' \right)
\label{eq:opt1}
\end{equation}
where the function $L_x$ is defined as $\sqrt{R^2 + x^2}$.
and $R$ is the radius of nanowire or nanotube. The potential is obtained to be:
\begin{multline}
\label{eq:opt1}
\frac{V^{1D}(x)}{V_0^{1D}} = sinh^{-1}(\frac{x}{R}) + sinh^{-1}(\frac{W_D-x}{R}) - sinh^{-1}(\frac{W_D}{R}) \\
+ \frac{ x L_x + (W_D-x) L_{W_D-x} - 2x(W_D-x) }{R^2} 
\end{multline}
where $V_0^{1D}$ is defined as $\frac{q N}{2 \epsilon} R^2$.
The depletion widths of 1D PN junctions exhibit an exponential dependence on $\Delta V \epsilon / N$
\begin{equation}
\label{eq:W1D}
W_D^{1D} \approx R ~ exp \left(\frac{2 \epsilon \Delta V}{q N R^2} - \frac{1}{2}\right)
\end{equation}
	
Figure 5b compares the results of analytical model and numerical simulation of 2D and 1D PN junctions with different thicknesses showing a good agreement. Moreover, it demonstrates the significant impact of the flake thickness and nanowire diameter on the potential profile. Increasing the thickness of low dimensional material enhances the depleted charge and electric field considering a constant volume doping density. 

To make the analytic model work for  \textcolor{black}{all thicknesses even beyond $W_D$, the simplifications used to approximate the depletion width equations (\ref{eq:W1D}) and (\ref{eq:W2D}) should be avoided. These simplifications are: a) Semiconductor thickness is much smaller than depletion width ($T \ll W_D$ and $R \ll W_D$), and b) Ignoring the E-field of distant charges (farther than $W_D$).} The exact analytic equations for potential profile in 2D and 1D are listed in table I. It is noteworthy that the potential attains an additional factor $\beta$ for large thicknesses due to the impact of distant charges as shown in table I. 
\textcolor{black}{$\beta$ equals 2 for thicknesses close to 0, whereas it decreases to 1 for infinitely thick junctions (equivalent to 3D).} 
Figure 6 shows the depletion width as a function of junction thickness in different dimensions. For a thickness beyond depletion width, the fringing field gets screened at the surface and $W_D$ gets close to the corresponding value in 3D. However, for smaller thicknesses, $W_D$ deviates from 3D case. Especially, 1D case shows a significantly higher $W_D$ values and increased sensitivity to the diameter. $W_D$ has a similar response to variations in $N$. The increased sensitivity of $W_D$ with respect to junction parameters in low dimensions enables the possibility of new sensors. 
  
\begin{figure}[!b]
        \centering
        \begin{subfigure}[b]{0.5\textwidth}
               \includegraphics[width=\textwidth]{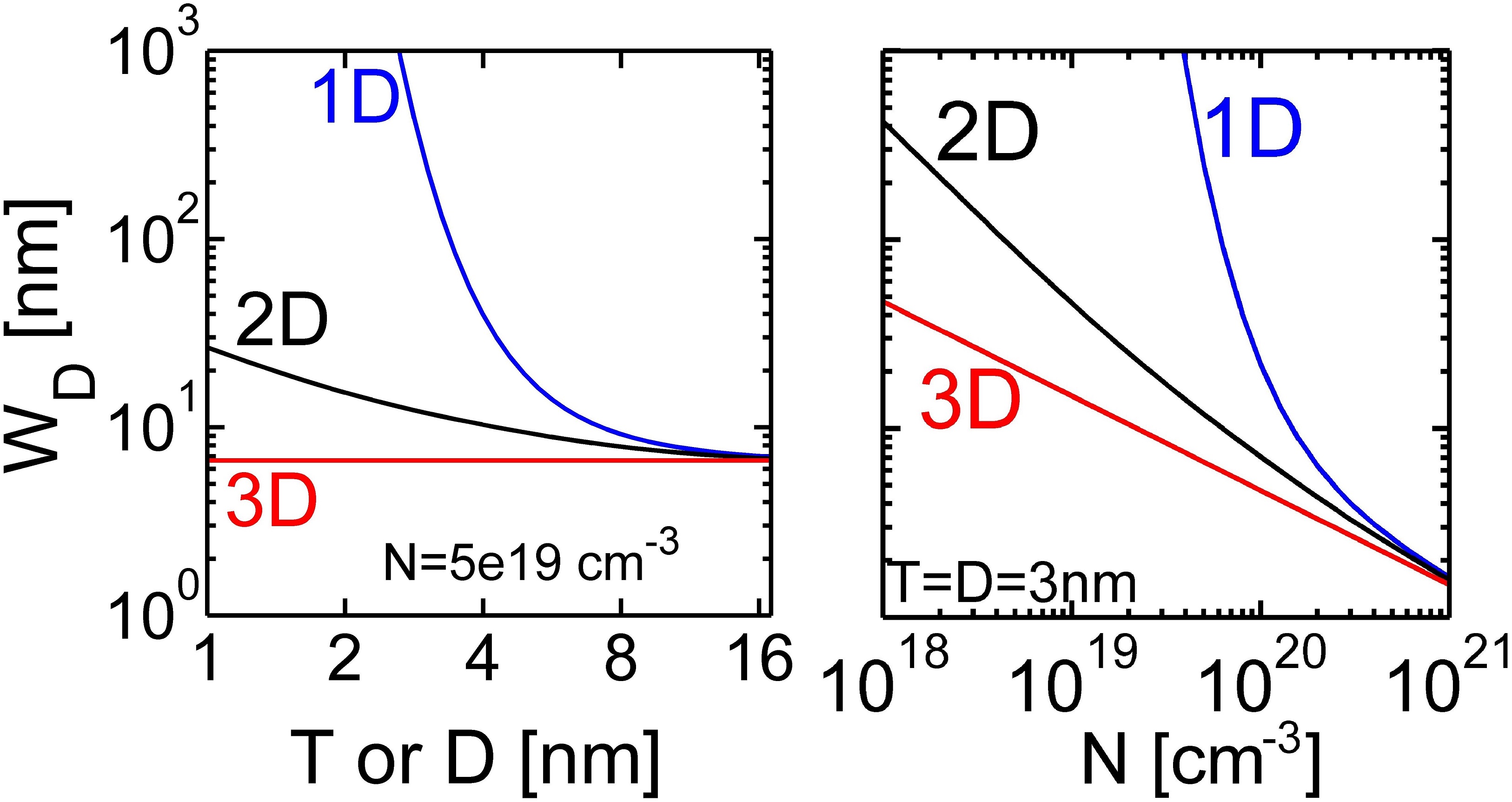}
               \vspace{+.3\baselineskip}
                \label{fig:Same_EoT}
        \end{subfigure}%
        \vspace{-1.5\baselineskip}        
        \caption{\textbf{ \textcolor{black}{Analytic} $W_D$ as a function of thickness (left) and N (right).} }\label{fig:Fig1}
\end{figure}
\begin{table}[!t]
               \vspace{+0.2\baselineskip}
\normalsize
\captionof{table}{ \textbf{V(x) in 2D and \textcolor{black}{1D} valid for all thicknesses.}}
               \vspace{-0.4\baselineskip}
\resizebox{0.99\textwidth}{!}{
\renewcommand{\arraystretch}{1.3}
\centering
\begin{tabular} { |c|c|c| }
  \hline
  V(x) & $ \frac{q N}{2 \epsilon} \left( \Gamma(W_D) - \Gamma(x) - \Gamma(W_D-x) \right) $ & $ 0 < x < \frac{W_D}{2}$\\ \hline  
  2D & $ \Gamma(x) = \frac{1}{\pi} \beta ~ atan(\frac{2x}{T}) ( \frac{3}{4 }T^2 - x^2 ) +  \frac{\beta}{2} x^2 +  \frac{1}{\pi} x T ln(\frac{T^2}{4} + x^2)$  & $\beta = \frac{3}{2} - \frac{1}{\pi} atan(ln(\frac{2T}{W_D}))$ \\ \hline
  1D & $ \Gamma(x) = -R^2 asinh( \frac{x}{R}) - \beta ~x \sqrt{x^2 + R^2} + \frac{\beta x}{2}(x- \frac{W_D}{2}) $ & $\beta = \frac{3}{2} - \frac{1}{\pi} atan(ln(\frac{2 R}{W_D}))$ \\ \hline  
\end{tabular}}
\label{Table}
\end{table} 

\section{Approximate Equations}
\textcolor{black}{
Table II shows the approximate equations for depletion widths ($W_D$) of PN junctions in different dimensions as obtained in this work. The approximate $W_D$ equations in 1D and 2D are valid when the thickness of the semiconductor is much smaller than $W_D$ ($R \ll W_D$ and $T \ll W_D$). An error less than 10\% can be achieved using these approximations if T $<$ $W_D$/7 in 2D or R $<W_D$ in 1D.
}
Interestingly, the square root dependence of $W_D$ on $\epsilon / N$ in 3D, changes to a linear and exponential dependence for 2D and 1D, respectively. Such an exponential dependence leads to high sensitivity of 1D PN junction to its control parameters, which can be used towards new sensors. For example, a small change in the biasing of 1D PN junction affects $W_D$ significantly through $\Delta V$. Such a junction under illumination can result in a \textcolor{black}{large} output current response. 
  
\begin{table}[!b]
        \centering
\vspace{1.7\baselineskip}
\captionof{table}{\textbf{Depletion widths of 1D, 2D and 3D PN junctions. \textcolor{black}{$R$ and $T$ are nanowire radius and flake thickness in 1D and 2D, respectively. $\epsilon$, $N$ and $\Delta V$ are the dielectric constant, doping density and potential drop across the PN junction.}}}             
\resizebox{0.46\textwidth}{!}{
\begin{tabular}{lr} 
$W_D$ & Dimension \\
\hline
$ 2 \sqrt{ \frac{\Delta V \epsilon}{q N} } $ & 3D  \\
$\frac{\pi \epsilon \Delta V}{ln(4) ~ q N T} $ & 2D \\
$ R ~ exp \left(\frac{2 \epsilon \Delta V}{q N R^2} - \frac{1}{2}\right)$ & 1D 
\end{tabular} }
\end{table} 

\section{Conclusion}
In summary, using a new analysis, the electrostatics of PN junctions in low dimensional material systems have been shown to differ significantly from the 3D junctions. Reducing the dimensionality increases the depletion width and its sensitivity to doping and thickness of PN junction, attractive towards sensing applications. The analytic results match closely with experimental measurements and numerical simulations.

\clearpage
\newpage

\bibliographystyle{ieeetr}
\bibliography{thesis}

\begin{thebibliography}{10}

\bibitem{Geim}
A.~K. Geim and K.~S. Novoselov, ``The rise of graphene,'' {\em Nature
  materials}, vol.~6, no.~3, pp.~183--191, 2007.

\bibitem{Dres}
M.~S. Dresselhaus, G.~Chen, M.~Y. Tang, R.~Yang, H.~Lee, D.~Wang, Z.~Ren, J.-P.
  Fleurial, and P.~Gogna, ``New directions for low-dimensional thermoelectric
  materials,'' {\em Advanced Materials}, vol.~19, no.~8, pp.~1043--1053, 2007.

\bibitem{MoS2_photo}
M.~S. Choi, D.~Qu, D.~Lee, X.~Liu, K.~Watanabe, T.~Taniguchi, and W.~J. Yoo,
  ``Lateral mos2 p--n junction formed by chemical doping for use in
  high-performance optoelectronics,'' {\em ACS nano}, vol.~8, no.~9,
  pp.~9332--9340, 2014.

\bibitem{FrankGR}
F.~Schwierz, ``Graphene transistors,'' {\em Nature nanotechnology}, vol.~5,
  no.~7, pp.~487--496, 2010.

\bibitem{CNT}
A.~D. Franklin, M.~Luisier, S.-J. Han, G.~Tulevski, C.~M. Breslin, L.~Gignac,
  M.~S. Lundstrom, and W.~Haensch, ``Sub-10 nm carbon nanotube transistor,''
  {\em Nano letters}, vol.~12, no.~2, pp.~758--762, 2012.

\bibitem{Fan1}
F.~W. Chen, H.~Ilatikhameneh, G.~Klimeck, Z.~Chen, and R.~Rahman,
  ``Configurable electrostatically doped high performance bilayer graphene
  tunnel fet,'' {\em IEEE Journal of the Electron Devices Society}, vol.~4,
  no.~3, pp.~124--128, 2016.

\bibitem{Fiori2}
B.~N. Szafranek, G.~Fiori, D.~Schall, D.~Neumaier, and H.~Kurz, ``Current
  saturation and voltage gain in bilayer graphene field effect transistors,''
  {\em Nano letters}, vol.~12, no.~3, pp.~1324--1328, 2012.

\bibitem{Fan2}
F.~W. Chen, H.~Ilatikhameneh, G.~Klimeck, R.~Rahman, T.~Chu, and Z.~Chen,
  ``Achieving a higher performance in bilayer graphene fet-strain
  engineering,'' in {\em 2015 International Conference on Simulation of
  Semiconductor Processes and Devices (SISPAD)}, pp.~177--181, IEEE, 2015.

\bibitem{MoS2}
B.~Radisavljevic, A.~Radenovic, J.~Brivio, i.~V. Giacometti, and A.~Kis,
  ``Single-layer mos2 transistors,'' {\em Nature nanotechnology}, vol.~6,
  no.~3, pp.~147--150, 2011.

\bibitem{Mehdi}
M.~Salmani-Jelodar, S.~R. Mehrotra, H.~Ilatikhameneh, and G.~Klimeck, ``Design
  guidelines for sub-12 nm nanowire mosfets,'' {\em IEEE Transactions on
  Nanotechnology}, vol.~14, no.~2, pp.~210--213, 2015.

\bibitem{Hesam1}
H.~Ilatikhameneh, Y.~Tan, B.~Novakovic, G.~Klimeck, R.~Rahman, and
  J.~Appenzeller, ``Tunnel field-effect transistors in 2-d transition metal
  dichalcogenide materials,'' {\em IEEE Journal on Exploratory Solid-State
  Computational Devices and Circuits}, vol.~1, pp.~12--18, 2015.

\bibitem{Sarkar}
D.~Sarkar, X.~Xie, W.~Liu, W.~Cao, J.~Kang, Y.~Gong, S.~Kraemer, P.~M. Ajayan,
  and K.~Banerjee, ``A subthermionic tunnel field-effect transistor with an
  atomically thin channel,'' {\em Nature}, vol.~526, no.~7571, pp.~91--95,
  2015.

\bibitem{Hesam3}
H.~Ilatikhameneh, T.~A. Ameen, G.~Klimeck, J.~Appenzeller, and R.~Rahman,
  ``Dielectric engineered tunnel field-effect transistor,'' {\em IEEE Electron
  Device Letters}, vol.~36, no.~10, pp.~1097--1100, 2015.

\bibitem{Fiori_Nature}
G.~Fiori, F.~Bonaccorso, G.~Iannaccone, T.~Palacios, D.~Neumaier, A.~Seabaugh,
  S.~K. Banerjee, and L.~Colombo, ``Electronics based on two-dimensional
  materials,'' {\em Nature nanotechnology}, vol.~9, no.~10, pp.~768--779, 2014.

\bibitem{Hesam2}
H.~Ilatikhameneh, G.~Klimeck, J.~Appenzeller, and R.~Rahman, ``Scaling theory
  of electrically doped 2d transistors,'' {\em IEEE Electron Device Letters},
  vol.~36, no.~7, pp.~726--728, 2015.

\bibitem{Tarek1}
T.~A. Ameen, H.~Ilatikhameneh, G.~Klimeck, and R.~Rahman, ``Few-layer
  phosphorene: An ideal 2d material for tunnel transistors,'' {\em Scientific
  Reports}, vol.~6, p.~28515, 2016.

\bibitem{Hesam4}
H.~Ilatikhameneh, T.~A. Ameen, B.~Novakovic, G.~Klimeck, and R.~Rahman,
  ``Saving moore's law down to 1 nm channels with anisotropic effective mass,''
  {\em Scientific Reports}, vol.~6, p.~31501, 2016.

\bibitem{Thermo}
A.~A. Balandin, ``Thermal properties of graphene and nanostructured carbon
  materials,'' {\em Nature materials}, vol.~10, no.~8, pp.~569--581, 2011.

\bibitem{Si_Thermo}
A.~I. Boukai, Y.~Bunimovich, J.~Tahir-Kheli, J.-K. Yu, W.~A. Goddard~Iii, and
  J.~R. Heath, ``Silicon nanowires as efficient thermoelectric materials,''
  {\em Nature}, vol.~451, no.~7175, pp.~168--171, 2008.

\bibitem{Si_Thermo2}
A.~I. Hochbaum, R.~Chen, R.~D. Delgado, W.~Liang, E.~C. Garnett, M.~Najarian,
  A.~Majumdar, and P.~Yang, ``Enhanced thermoelectric performance of rough
  silicon nanowires,'' {\em Nature}, vol.~451, no.~7175, pp.~163--167, 2008.

\bibitem{MoS2_photo2}
Y.~Deng, Z.~Luo, N.~J. Conrad, H.~Liu, Y.~Gong, S.~Najmaei, P.~M. Ajayan,
  J.~Lou, X.~Xu, and P.~D. Ye, ``Black phosphorus--monolayer mos2 van der waals
  heterojunction p--n diode,'' {\em ACS nano}, vol.~8, no.~8, pp.~8292--8299,
  2014.

\bibitem{Tao}
T.~Chu, H.~Ilatikhameneh, G.~Klimeck, R.~Rahman, and Z.~Chen, ``Electrically
  tunable bandgaps in bilayer mos2,'' {\em Nano letters}, vol.~15, no.~12,
  pp.~8000--8007, 2015.

\bibitem{Kane1}
E.~O. Kane, ``Theory of tunneling,'' {\em Journal of Applied Physics}, vol.~32,
  no.~1, pp.~83--91, 1961.

\bibitem{SB1}
A.~Seabaugh, ``Promise of tunnel diode integrated circuits,'' in {\em Tunnel
  Diode and CMOS/HBT Integration Workshop}, pp.~1--13, 1999.

\bibitem{Analytic2}
H.~Ilatikhameneh, R.~B. Salazar, G.~Klimeck, R.~Rahman, and J.~Appenzeller,
  ``From fowler nordheim to nonequilibrium green's function modeling of
  tunneling,'' {\em IEEE Transactions on Electron Devices}, vol.~63,
  pp.~2871--2878, July 2016.

\bibitem{App1}
J.~Appenzeller, Y.-M. Lin, J.~Knoch, and P.~Avouris, ``Band-to-band tunneling
  in carbon nanotube field-effect transistors,'' {\em Physical review letters},
  vol.~93, no.~19, p.~196805, 2004.

\bibitem{Ionescu}
A.~M. Ionescu and H.~Riel, ``Tunnel field-effect transistors as
  energy-efficient electronic switches,'' {\em Nature}, vol.~479, no.~7373,
  pp.~329--337, 2011.

\bibitem{Sub10}
H.~Ilatikhameneh, G.~Klimeck, and R.~Rahman, ``Can homojunction tunnel fets
  scale below 10 nm?,'' {\em IEEE Electron Device Letters}, vol.~37, no.~1,
  pp.~115--118, 2016.

\bibitem{Analytic1}
R.~B. Salazar, H.~Ilatikhameneh, R.~Rahman, G.~Klimeck, and J.~Appenzeller, ``A
  predictive analytic model for high-performance tunneling field-effect
  transistors approaching non-equilibrium green's function simulations,'' {\em
  Journal of Applied Physics}, vol.~118, no.~16, p.~164305, 2015.

\bibitem{Exp2D}
D.~Reuter, C.~Werner, A.~Wieck, and S.~Petrosyan, ``Depletion characteristics
  of two-dimensional lateral p-n-junctions,'' {\em Applied Physics Letters},
  vol.~86, no.~16, 2005.

\bibitem{Changxi}
C.~Zheng, Q.~Zhang, B.~Weber, H.~Ilatikhameneh, F.~Chen, H.~Sahasrabudhe,
  R.~Rahman, S.~Li, Z.~Chen, J.~Hellerstedt, {\em et~al.}, ``Direct observation
  of 2d electrostatics and ohmic contacts in template-grown graphene/ws2
  heterostructures.,'' {\em ACS nano}, 2017.

\bibitem{Leonard}
F.~L{\'e}onard and J.~Tersoff, ``Novel length scales in nanotube devices,''
  {\em Physical Review Letters}, vol.~83, no.~24, p.~5174, 1999.

\bibitem{2d_theo_1}
A.~S. Achoyan, A.~Yesayan, E.~Kazaryan, and S.~Petrosyan, ``Two-dimensional pn
  junction under equilibrium conditions,'' {\em Semiconductors}, vol.~36,
  no.~8, pp.~903--907, 2002.

\bibitem{Khor1}
B.~Gharekhanlou and S.~Khorasani, ``Current--voltage characteristics of
  graphane pn junctions,'' {\em IEEE Transactions on Electron Devices},
  vol.~57, no.~1, pp.~209--214, 2010.

\bibitem{Khor2}
B.~Gharekhanlou, S.~Khorasani, and R.~Sarvari, ``Two-dimensional bipolar
  junction transistors,'' {\em Materials Research Express}, vol.~1, no.~1,
  p.~015604, 2014.

\bibitem{2d_theo_2}
H.~Yu, A.~Kutana, and B.~I. Yakobson, ``Carrier delocalization in
  two-dimensional coplanar p--n junctions of graphene and metal
  dichalcogenides,'' {\em Nano Letters}, vol.~16, no.~8, pp.~5032--5036, 2016.

\bibitem{Exp1D}
F.~L{\'e}onard and J.~Tersoff, ``Novel length scales in nanotube devices,''
  {\em Physical Review Letters}, vol.~83, no.~24, p.~5174, 1999.

\bibitem{Doping1}
L.~Yang, K.~Majumdar, H.~Liu, Y.~Du, H.~Wu, M.~Hatzistergos, P.~Hung,
  R.~Tieckelmann, W.~Tsai, C.~Hobbs, {\em et~al.}, ``Chloride molecular doping
  technique on 2d materials: Ws2 and mos2,'' {\em Nano letters}, vol.~14,
  no.~11, pp.~6275--6280, 2014.

\bibitem{Doping2}
A.~Azcatl, X.~Qin, A.~Prakash, C.~Zhang, L.~Cheng, Q.~Wang, N.~Lu, M.~J. Kim,
  J.~Kim, K.~Cho, {\em et~al.}, ``Covalent nitrogen doping and compressive
  strain in mos2 by remote n2 plasma exposure,'' {\em Nano Letters}, vol.~16,
  no.~9, pp.~5437--5443, 2016.

\bibitem{Doping3}
P.~Ayala, R.~Arenal, M.~R{\"u}mmeli, A.~Rubio, and T.~Pichler, ``The doping of
  carbon nanotubes with nitrogen and their potential applications,'' {\em
  Carbon}, vol.~48, no.~3, pp.~575--586, 2010.

\bibitem{Mehdi1}
M.~Salmani-Jelodar, S.~R. Mehrotra, H.~Ilatikhameneh, and G.~Klimeck, ``Design
  guidelines for sub-12 nm nanowire mosfets,'' {\em IEEE Transactions on
  Nanotechnology}, vol.~14, no.~2, pp.~210--213, 2015.

\bibitem{Mehdi2}
M.~S. Jelodar, H.~Ilatikhameneh, P.~Sarangapani, S.~R. Mehrotra, G.~Klimeck,
  S.~Kim, and K.~Ng, ``Tunneling: The major issue in ultra-scaled mosfets,'' in
  {\em Nanotechnology (IEEE-NANO), 2015 IEEE 15th International Conference on},
  pp.~670--673, IEEE, 2015.

\bibitem{Mehdi3}
M.~Salmani-Jelodar, H.~Ilatikhameneh, S.~Kim, K.~Ng, P.~Sarangapani, and
  G.~Klimeck, ``Optimum high-k oxide for the best performance of ultra-scaled
  double-gate mosfets,'' {\em IEEE Transactions on Nanotechnology}, vol.~15,
  no.~6, pp.~904--910, 2016.

\bibitem{N5}
J.~E. Fonseca, T.~Kubis, M.~Povolotskyi, B.~Novakovic, A.~Ajoy, G.~Hegde,
  H.~Ilatikhameneh, Z.~Jiang, P.~Sengupta, Y.~Tan, {\em et~al.}, ``Efficient
  and realistic device modeling from atomic detail to the nanoscale,'' {\em
  Journal of Computational Electronics}, vol.~12, no.~4, pp.~592--600, 2013.

\bibitem{N5_2}
J.~Sellier, J.~Fonseca, T.~C. Kubis, M.~Povolotskyi, Y.~He, H.~Ilatikhameneh,
  Z.~Jiang, S.~Kim, D.~Mejia, P.~Sengupta, {\em et~al.}, ``Nemo5, a parallel,
  multiscale, multiphysics nanoelectronics modeling tool,'' in {\em Proc.
  SISPAD}, pp.~1--4, 2012.

\end{thebibliography}

\end{document}